\documentclass[12pt]{article} 
\usepackage[koi8-r]{inputenc}
\usepackage[english]{babel}
\usepackage{pazh}
\usepackage{graphicx,amssymb,amsmath}
\usepackage{natbib}
\usepackage{xspace}
\tightenlines

\newcommand{\Teff}{T_{\rm eff}}
\newcommand{\logg}{\rm log~ g}
\newcommand{\kms}{km\,s$^{-1}$}
\newcommand{\eps}[1]{\log\varepsilon_{\rm #1}}
\newcommand{\Eexc}{$E_{\rm exc}$}
\newcommand{\eu}[5]{\mbox{$#1\,^#2{\rm #3}^{#4}_{\rm #5}$}}

\def\arxivprefixesep{:}

\newcommand{\eprint}[2][]{%
	{\tt\if!#1!#2\else#1\arxivprefixesep\ignorespaces#2\fi}%
}

\sloppypar

\begin{document}
	
	\baselineskip 21pt
	
	
	\title{\bf Non-LTE analysis of the Si~II lines in $\iota$~Her \\ with various atomic data sets}
	
	\author{\bf \hspace{-1.3cm}\copyright\, 2022 \ \
		L.~Mashonkina\affilmark{1*},
		T.~Sitnova\affilmark{1**},
		S.~Korotin\affilmark{2***}
	}
	\affil{
		{\it Institute of Astronomy, Russian Academy of Sciences, Pyatnitskaya st. 48, 119017 Moscow, Russia}$^1$\\
		Crimean Astrophysical Observatory, Nauchny 298409, Republic of Crimea$^2$
	}
	
	\vspace{2mm}
	
	\sloppypar 
	\vspace{2mm}

This study shows that the statistical equilibrium of Si~II in the atmosphere of a B3~IV type star $\iota$~Her is extremely sensitive to a variation in photoionization cross-sections for the Si~II levels. The difference in abundances derived from absorption lines of Si~II between applying the data from two equal accuracy sources, namely, the Opacity Project (OP) and the NORAD database, amounts to 0.18~dex, on average. Using the hydrogenic approximation for photoionization cross-sections, we obtain the departure coefficients for the Si~II \eu{4s}{2}{S}{}{} level, the source function for Si~II 6371~\AA, and the abundance derived from this line, which are very similar to the corresponding values computed by Takeda (2022). We suppose that close-to-solar abundance obtained by Takeda (2022) from Si~II 6371~\AA\ in $\iota$~Her is due to using the hydrogenic photoionization cross-sections for the Si~II levels. However, emission lines of Si~II observed in $\iota$~Her can only be reproduced with the OP photoionization cross-sections. Photoionization cross-sections for the Si~II levels need further improvements.


\noindent

\vfill
\noindent\rule{8cm}{1pt}\\
{$^*$ e-mail $<$lima@inasan.ru$>$}\\
{$^{**}$ e-mail $<$sitnova@inasan.ru$>$}\\
{$^{***}$ e-mail $<$serkor1@mail.ru$>$}

\section{Introduction}

The star HD~160762 ($\iota$~Her) is a B3~IV-type, chemically normal and slowly rotating star, with $v\sin i$ = 6\,\kms\ \citep{NP2012}. Its red-region spectrum reveals weak emission lines of various chemical species, as registered by Alexeeva et~al. (2016, C~I), Sadakane \& Nishimura (2017, Si~II), Sitnova et~al. (2018a, Ca~II), Sitnova et~al. (2018b, Fe~II), and Sadakane \& Nishimura (2019, C~I, C~II, N~I, Al~II, Si~II, Ca~II, Cr~II, Mn~II, Fe~II, Ni~II). Modelling of the line formation in a classical hydrostatic model atmosphere under the assumption of local thermodynamical equilibrium (LTE) can only yield an absorption line. An emission line in the star's spectrum is a signature of either the departures from LTE (non-LTE = NLTE effects) or/and a non-fulfillment of the model atmosphere assumptions. Alexeeva et~al. (2016), Sitnova et~al. (2018a), and Mashonkina (2020) have shown that, in case of $\iota$~Her, the NLTE effects are responsible for emission lines of C~I, Ca~II, and Si~II, respectively.

Having success with reproducing 10 emission lines of Si~II in $\iota$~Her, Mashonkina (2020, hereafter M2020) failed to achieve consistent abundances from different absorption lines. While the NLTE abundances from the Si~III lines were found to be solar, the NLTE abundances from the Si~II 6347, 6371\,\AA\ lines are substantially higher, by 0.8~dex.
Recently, Takeda (2022, hereafter T2021) constructed his own model atom of Si~I-II-III aiming to analyse exclusively the Si~II 6347, 6371\,\AA\ lines in a sample of late A through late B dwarfs. For the comparison with M2020, T2021 extended his NLTE calculations to an effective temperature of $\Teff$ = 20\,000~K. In contrast to M2020, who computed the NLTE abundance correction $\Delta_{\rm NLTE} = \eps{NLTE}-\eps{LTE}$ = 0.67~dex for Si~II 6371\,\AA\ in $\iota$~Her, T2021 obtained a negative NLTE correction of $\Delta_{\rm NLTE} = -0.1$~dex for $\Teff$ = 17\,500~K corresponding to $\iota$~Her. With such a NLTE correction, Si~II 6371\,\AA\ yields close-to-solar Si abundance of $\iota$~Her. T2021 has supposed that M2020 underestimated background opacity in the ultraviolet (UV) spectral range, resulting in an overestimated intensity of the UV ionizing radiation, enhanced overionization of Si~II, and overestimated positive $\Delta_{\rm NLTE}$ for lines of Si~II. 

This paper aims to investigate an influence of varying the atomic data used in calculations of the  statistical equilibrium (SE) of Si~I-II-III on the NLTE profiles of the Si~II lines and derived NLTE abundances and make clear the source of discrepancies between M2020 and T2021. 
Section~\ref{sect:method} describes a method of our calculations, including five different model atoms for Si~I-II-III. The computed NLTE effects are discussed in Sect.~\ref{sect:nlte}, with focusing on that for Si~II. Section~\ref{sect:conclusion} summarises our conclusions. 

\section{Method of calculations}\label{sect:method}

The observed spectrum was taken from Common Archive Observation data base\footnote{http://www.cfht.hawaii.edu/Instruments/Spectroscopy/Espadons/}. It was obtained with a spectral resolving power of R = $\lambda/\Delta\lambda \simeq$ 65\,000, using the ESPaDOnS instrument of the Canada-France-Hawaii Telescope (CHFT). Following M2020, we employ atmospheric parameters of $\iota$~Her, as derived by \citet{NP2012}: $\Teff$ = 17\,500~K, surface gravity with $\logg$ = 3.8, [Fe/H] = 0.02, microturbulent velocity of $\xi_t$ = 1~\kms, and the classical plane-parallel and LTE model atmosphere calculated with the code \textsc{LLmodels} (Shulyak et al. 2004). The Si~II and Si~III lines used in detailed analysis are listed in Table~\ref{Tab:abund}. We apply the same line atomic parameters as in Mashonkina (2020). We use the abundance scale where $\eps{H}$ = 12 and adopt the solar system silicon abundance, $\eps{\odot}$ = 7.51$\pm$0.01, from Lodders (2021).

\subsection{Codes}

We use the revised code {\sc detail} \citep{giddings81,butler84} to solve the coupled radiation transfer and statistical equilibrium (SE) equations. The updates were implemented by Keith Butler (Munich University, not published). The most important of them concern with 
\begin{itemize}
	\item implementing the accelerated lambda iteration (ALI) scheme following the efficient method of \citet{RH91,RH92},
	\item using the Opacity Project (OP, see Seaton et al. 1994, for a general review) photoionization cross-sections for the calculations of b-f absorption of C~I, N~I, Mg~I, Si~I, Al~I, and Fe~I in the opacity package,
	\item accounting for the line opacity introduced by H~I and metal lines by explicitly including
	it in solving the radiation transfer. The metal line list was extracted from the Kurucz (1994) compilation.
\end{itemize}

\begin{figure}  
	\centering
	\includegraphics[width=0.6\columnwidth,clip]{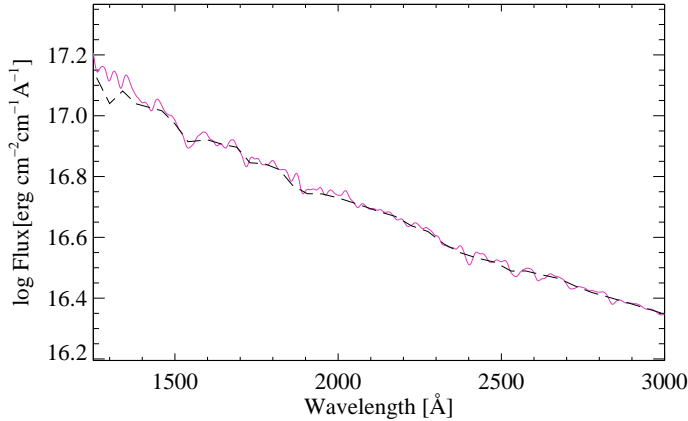}
	\caption{Emergent fluxes in the model of $\Teff$ = 17500~K, $\logg$ = 3.8, and [Fe/H] = 0.02 calculated with the {\sc detail} (continuous curve) and {\sc atlas9} (dashed curve). For clearer illustration, the {\sc detail} fluxes were convolved with a Gaussian profile of $\Delta \lambda_{\rm D}$ = 10~\AA. }
	\label{fig:flux}
\end{figure}

\citet{Takeda2021} supposes that the source of discrepancies in the NLTE results between his and M2020 studies is a treatment of the background opacity. \citet{Takeda2021} uses the code described by \citet{Takeda91}. It implements the opacities included in the {\sc atlas9} program (Kurucz, 1993a) along with Kurucz's (1993b) line opacity distribution function (ODF). For a comparison with T2021, we calculated the emergent fluxes of the $\Teff$/$\logg$ = 17\,500/3.80 model using the {\sc atlas9} code and Kurucz's (1993b) ODF. Figure~\ref{fig:flux} compares the {\sc detail} and {\sc atlas9} fluxes in the 1250-3000~\AA\ range, where the ionization thresholds of the Si~II \eu{3p^2}{2}{D}{}{} ($\lambda_{\rm thr} \simeq$ 1307~\AA), \eu{4s}{2}{S}{}{} (1507~\AA), and \eu{3d}{2}{D}{}{} (1905~\AA) levels are located. Overionization of exactly these three levels caused by superthermal radiation of a non-local origin below their thresholds is the main driver of overall overionization of Si~II. It is evident that different techniques of line opacity evaluation lead to fairly consistent intensities of ionizing radiation and cannot be the source of discrepancies between M2020 and T2021.

Analysis of line profiles in the observed spectrum and abundance determinations are based in this study on line profile fitting. The theoretical NLTE and LTE spectra were computed with the code {\sc Synth}V\_NLTE (Tsymbal et al. 2019), which implements the pre-computed departure coefficients from the {\sc detail} code. 
The best fit to the observed spectrum was obtained automatically using the {\sc IDL binmag} code by \citet{Kochukhov_binmag}. 

\subsection{Model atoms}

Here, we use the model atom of Si~I-II-III developed by M2020 as our standard model atom. Hereafter, it is referred to as Si2op. This is a comprehensive model atom based on the most up-to-date atomic data availabe so far. The energy levels were taken from National Institute of Standards and Technology (NIST) database\footnote{https://physics.nist.gov/PhysRefData/ASD} (Kramida et al. 2019), the transition probabilities from calculations of R.~Kurucz\footnote{http://kurucz.harvard.edu/atoms/},  the photoionization cross-sections from the Opacity Project (OP) that are accessible in the TOPbase\footnote{cdsweb.u-strasbg.fr/cgi-bin/topbase/} database (Cunto et al. 1993). For Si~II that is in a focus of this study, the OP data are available for all the levels with an excitation energy of \Eexc\ $\le$ 15.5~eV, up to \eu{9p}{2}{P}{\circ}{}. The situation with electron-impact excitation data is not perfect. Nevertheless, the R-matrix calculations are available for 414 transitions in Si~II (Aggarwal and Keenan, 2014) and 210 transitions in Si~III \citep{coll_si3}. The semi-empirical formulae were applied for the remaining bound-bound (b-b) and bound-free (b-f) transitions.

Takeda (2022) developed the model atom of Si~I-II-III, with using the same OP photoionization cross-sections and collisional data. However, his model atom has the following simplifications compared with our model atom Si2op.

\begin{itemize}
	\item Si~I: energy levels up to \Eexc\ = 7.30 eV are included by T2021, while up to \Eexc\ = 8.14 eV in Si2op. The ionization threshold is $\chi_{\rm thr}$(Si~I) = 8.187 eV.
	\item Si~II: energy levels up to \Eexc\ = 15.254 eV and 15.895 eV are included in T2021 and Si2op, respectively. The ionization threshold is $\chi_{\rm thr}$(Si~II) = 16.346 eV.
	\item Si~III: energy levels up to \Eexc\ = 21.73 eV and 33.19 eV are included in T2021 and Si2op, respectively. The ionization threshold is $\chi_{\rm thr}$(Si~III) = 33.49 eV.
	\item T2021 uses the OP photoionization cross-sections for 10 lowest terms (\Eexc\ $<$ 6 eV) of Si~I and 10 lowest terms (\Eexc\ $<$ 12.6 eV) of Si~II, in contrast to 55 (\Eexc\ $<$ 7.64 eV) and 41 (\Eexc\ $<$ 15.6 eV) terms, respectively, in Si2op. Both T2021 and M2020 use the hydrogenic approximation for the remaining levels. 
\end{itemize} 

For the comparison with T2021, our model atom Si2op was transformed, such to include the Si~I levels up to \Eexc\ = 7.30 eV, the Si~II levels up to \Eexc\ = 15.254 eV, and the OP photoionization cross-sections for 10 lowest terms of Si~I and 10 lowest terms of Si~II. Hereafter, it is referred to as Si2takeda.

\begin{figure*}  
	\centering
	\includegraphics[width=0.3\columnwidth,clip]{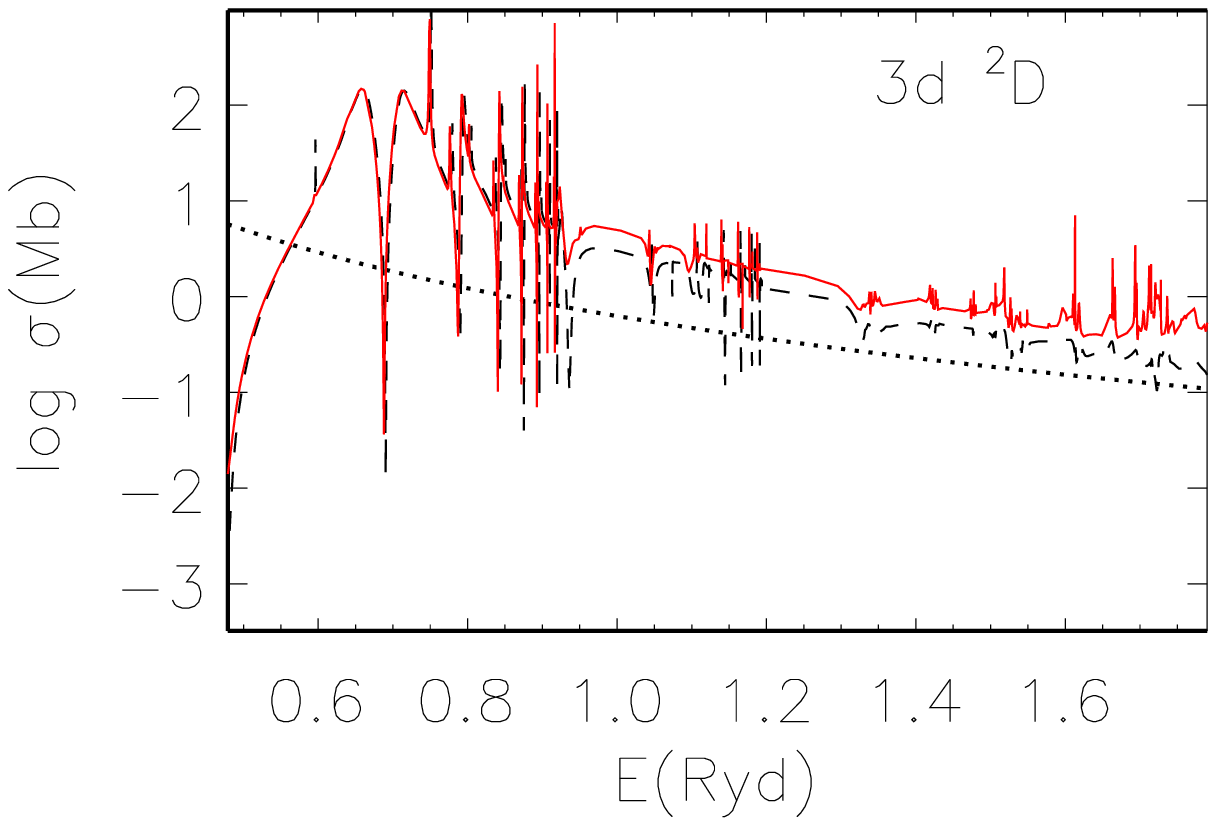}
	\includegraphics[width=0.3\columnwidth,clip]{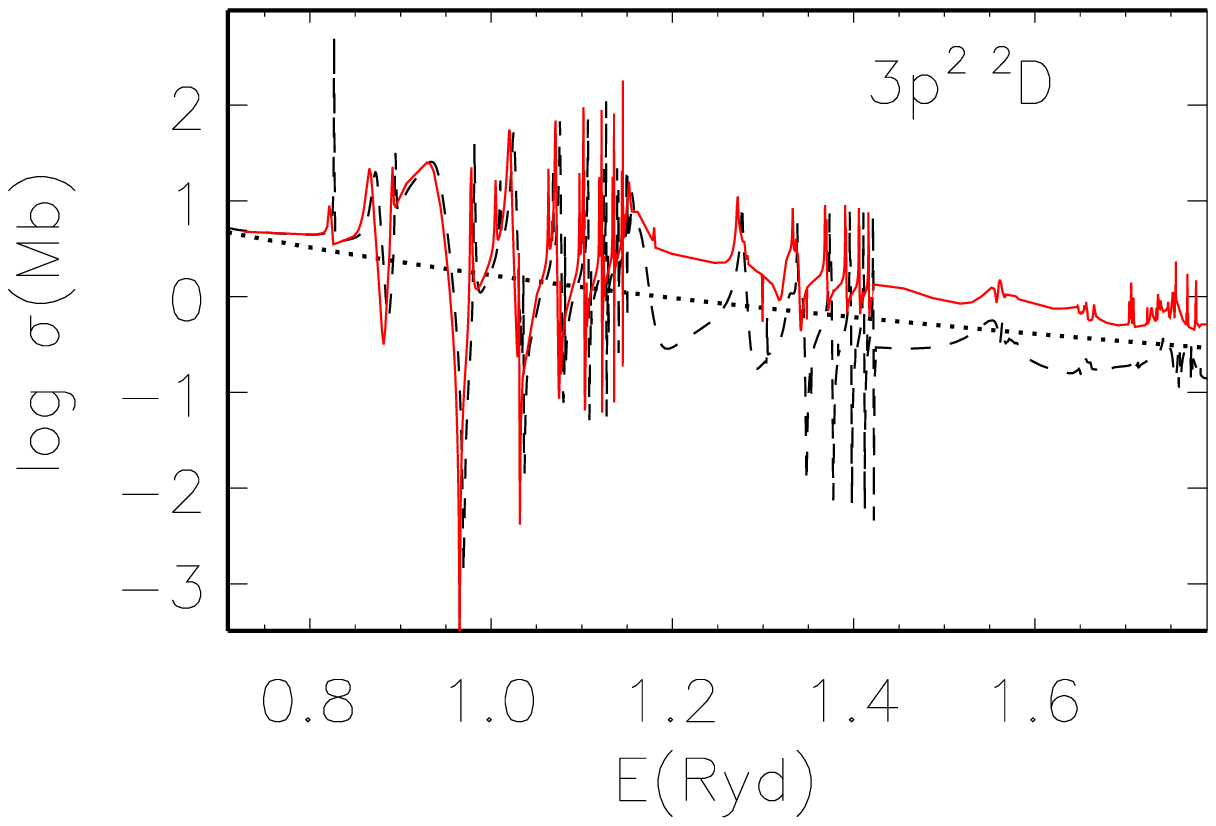}
	\includegraphics[width=0.3\columnwidth,clip]{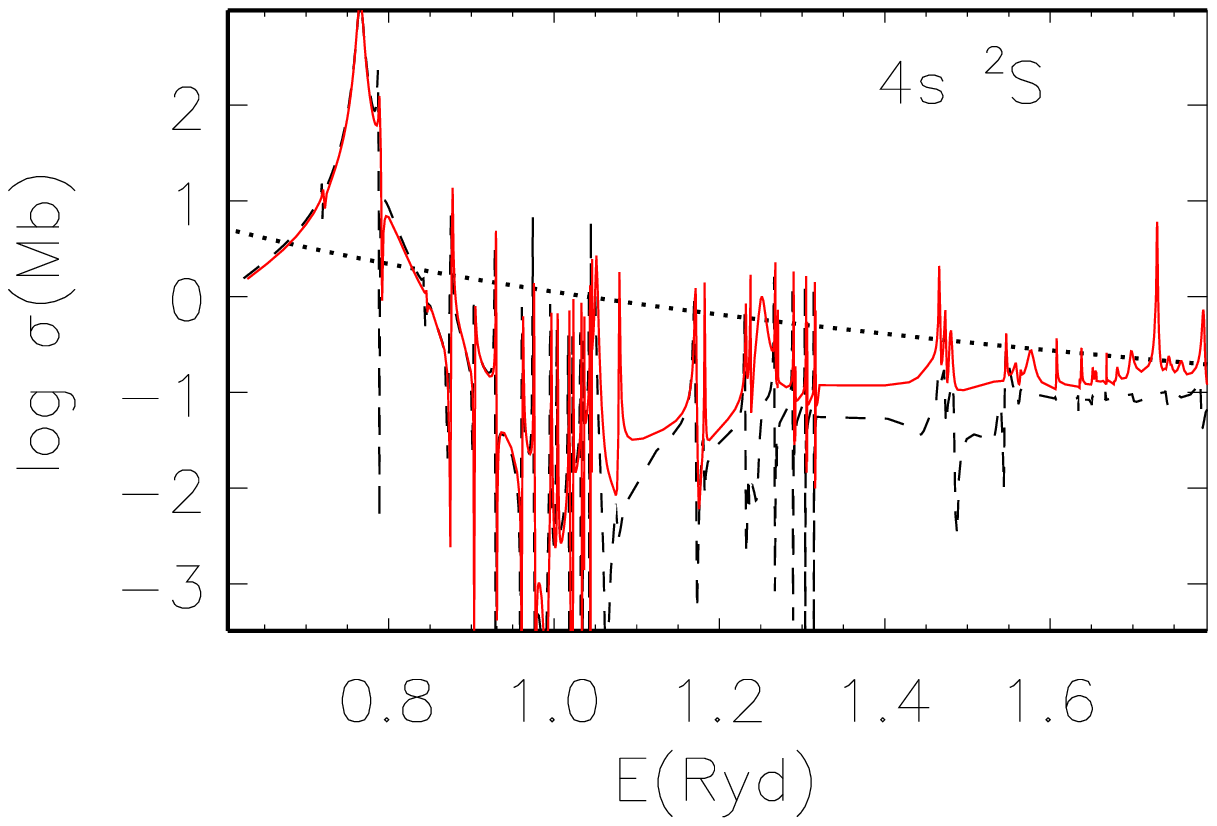}
	\caption{OP (solid curves) and NORAD (dashed curves) photoionization cross-sections for the Si~II \eu{3d}{2}{D}{}{}, \eu{3p^2}{2}{D}{}{}, and \eu{4s}{2}{S}{}{} levels as a function of photon energy. The hydrogenic approximation is shown by dotted curves.}
	\label{fig:sigma}
\end{figure*}

Based on Si2op, we produced another three model atoms. 

The model atom Si2hyd3d4s uses the hydrogenic photoionization cross-sections instead of the OP ones for the three selected levels, namely, Si~II \eu{4s}{2}{S}{}{}, \eu{3d}{2}{D}{}{}, and \eu{3p^2}{2}{D}{}{}. In the hydrogenic approximation, we replace a principal quantum number $n$ with an effective principal quantum number $n_{\rm eff}$. This model atom was already used in test calculations of M2020 (Figs. 5 and 6 in that paper).

The model atom Si2hyd is the same as Si2op except for the hydrogenic photoionization cross-sections instead of the OP ones for all the Si~II levels.

The model atom Si2norad employs the photoionization cross-sections computed by Nahar (1995) for Si~II, as available in the NORAD\footnote{https://norad.astronomy.osu.edu/} database. The NORAD data are available for the doublet terms up to \eu{9p}{2}{P}{\circ}{}. The OP data are used for Si~I and Si~III. Figure~\ref{fig:sigma} displays the photoionization cross-sections from Nahar (1995) and OP for the three atomic levels, which play an important role in the SE of Si~II.

\section{Non-LTE effects for Si~II in different line-formation scenarios}\label{sect:nlte}

We performed the NLTE calculations with the model atoms Si2takeda, Si2norad, and Si2hyd, analysed profiles of the Si~II lines in $\iota$~Her, including the emission lines, and determined the Si abundances from individual lines. We note that variations in a number of included levels and atomic data for Si~II do not influence on the Si~III level populations because Si~III is the majority species in atmosphere of $\iota$~Her. The LTE and NLTE results corresponding to the model atoms Si2op and Si2hyd3d4s were taken from M2020. Everywhere, the model atmosphere 17500/3.80/0.02 and $\xi_t$ = 1~\kms\ were employed.

\underline{Si2takeda versus Si2op}. We find that slight reducing total number of the Si~II levels compared with that in Si2op and replacing the OP photoionization cross-sections of the Si~II high-excitation levels with the hydrogenic ones cannot be the source of discrepancies between T2021 and 2020. With Si2takeda, we obtained only minor changes in the NLTE populations of the Si~II \eu{4s}{2}{S}{}{} and \eu{4p}{2}{P}{\circ}{} levels compared with those for Si2op. The abundance derived from the Si~II 6371~\AA\ line, which arises between these two levels, changes by less than 0.01~dex compared with that for Si2op. 

\underline{Si2hyd3d4s versus Si2op}. \citet{Mashonkina2020} noted that using the model atom Si2hyd3d4s leads to strengthened Si~II 6371~\AA\ line compared with the LTE case (Fig.~5 in M2020) and $\Delta_{\rm NLTE} =-0.14$~dex. Negative NLTE abundance correction for Si~II 6371~\AA, this is what was calculated by T2021. However, no emission line of Si~II in $\iota$~Her can be reproduced with Si2hyd3d4s (see Fig.~6 in M2020 for Si~II 6239~\AA). T2021 did not check emission lines of Si~II.

\begin{figure*}  
	\centering
	\includegraphics[width=0.45\columnwidth,clip]{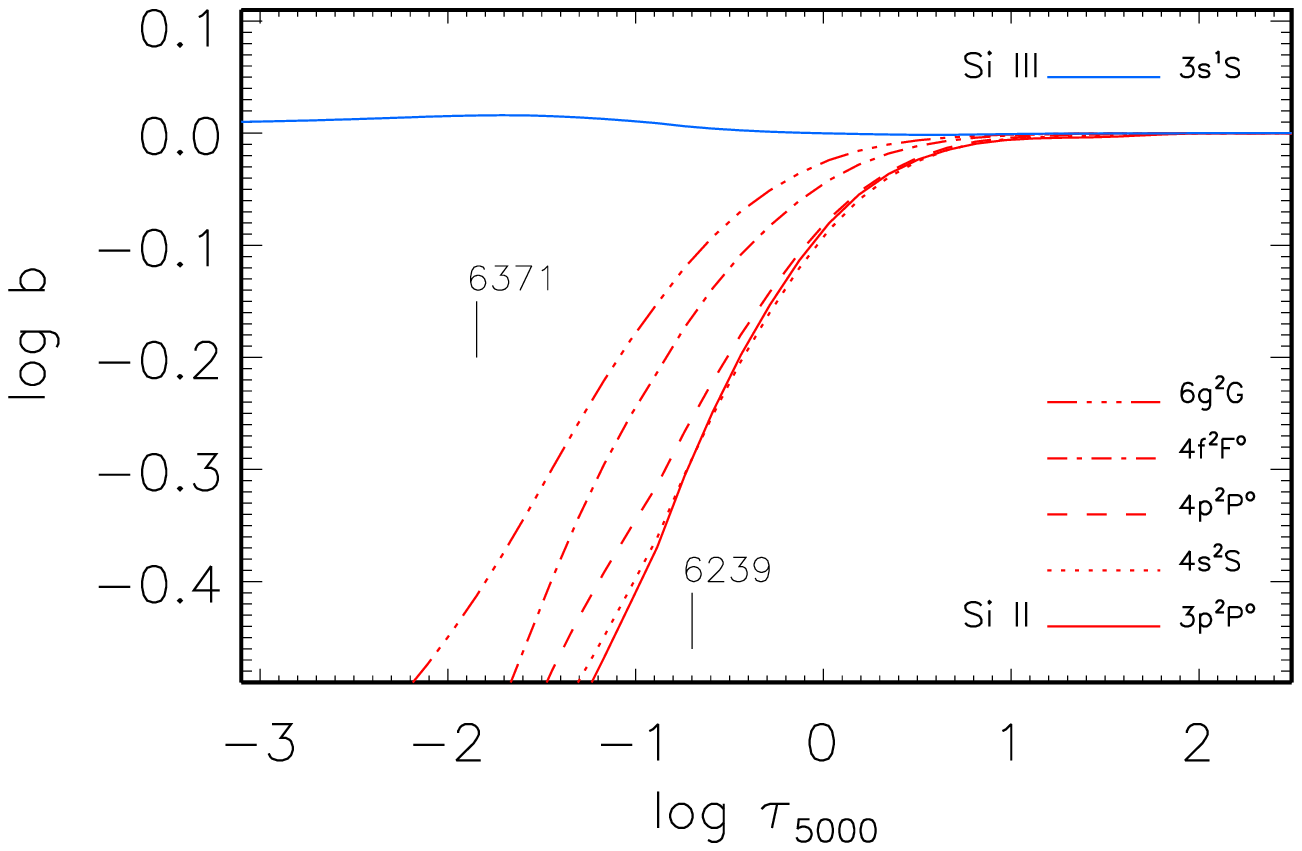}
	\includegraphics[width=0.45\columnwidth,clip]{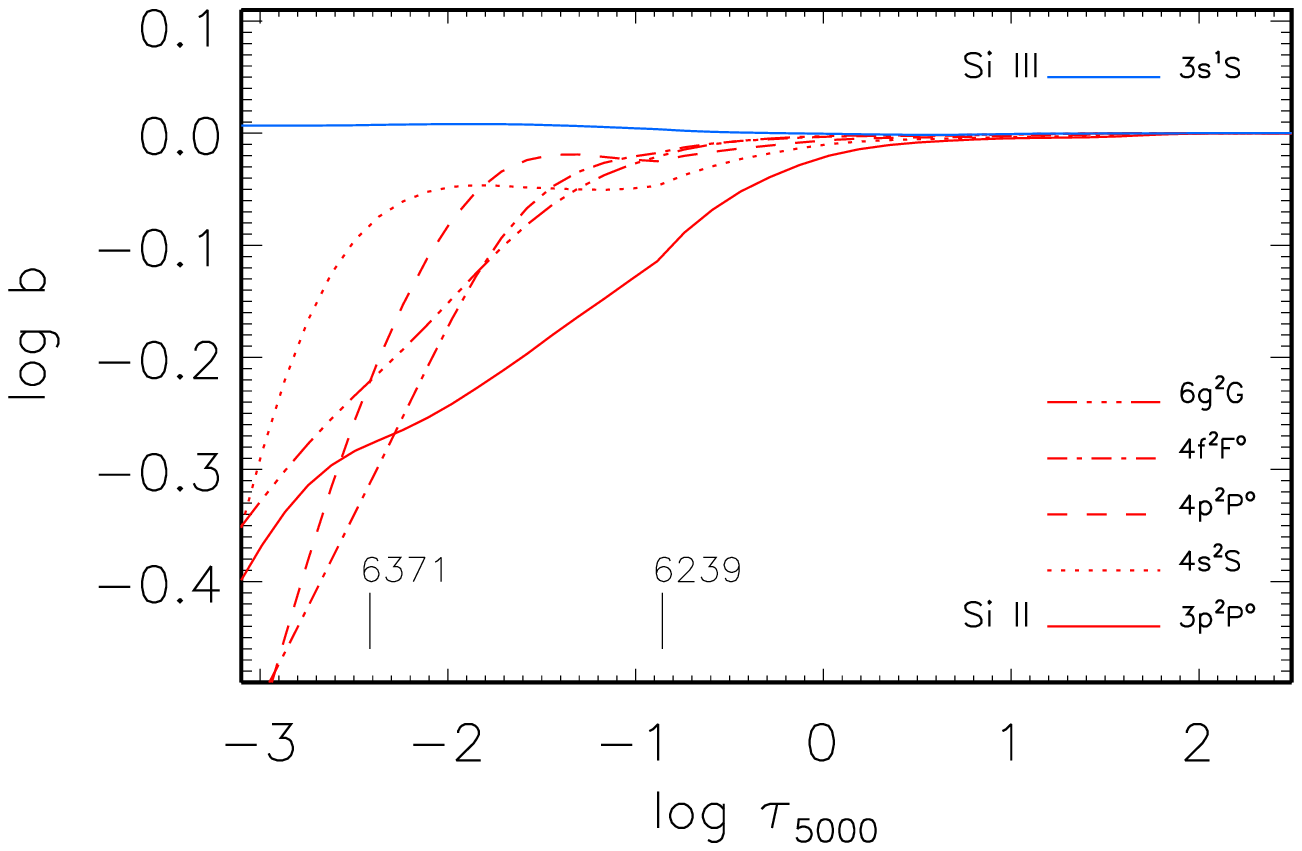}
	\caption{Departure coefficients, log~b, for the selected levels of Si~II as a function of log~$\tau_{5000}$ in the model atmosphere 17500/3.80/0.02 from calculations with the model atoms Si2op (left panel) and Si2hyd (right panel). Tick marks indicate the locations of line centre optical depth unity for Si~II 6371 and 6239~\AA.}
	\label{fig:bf}
\end{figure*}

\underline{Si2hyd versus Si2op}. Using the hydrogenic photoionization cross-sections for all the  Si~II levels produces great changes in the level populations, the line source functions, and derived Si abundances compared with those for the Si2op case.
Figure~\ref{fig:bf} displays the departure coefficients, ${\rm b = n_{NLTE}/n_{LTE}}$, for the selected Si~II levels, which are important for understanding the departures from LTE for Si~II 6371~\AA\ and 6239~\AA\ (\eu{4f}{2}{F}{\circ}{} - \eu{6g}{2}{G}{}{}). Here, ${\rm n_{NLTE}}$ and ${\rm n_{LTE}}$ are the statistical equilibrium and thermal (Saha-Boltzmann) number densities, respectively. The calculations were performed with Si2op and Si2hyd. In both cases, Si~II is subject to overionization starting from deep atmospheric layers, with log~$\tau_{5000} \sim 1$. However, in case of Si2hyd, the Si~II levels, in particular, \eu{4s}{2}{S}{}{} and \eu{4p}{2}{P}{\circ}{}, are depopulated to a less extent than in case of Si2op. 

With $b_{\rm low} < 1$ for the lower level of a transition, the corresponding line tends to become weaker. However, departures of the line strength from the LTE strength depend also on deviations of the line source function

$$ S_{\nu}^{l} =  \frac{2 h \nu^3}{c^2}\frac{1}{\frac{g_{\rm up}}{g_{\rm low}}\frac{n_{\rm low}}{n_{\rm up}} - 1} =
B_\nu(T) \frac{e^{h\nu/k T} -1}{b_{\rm low}/b_{\rm up} e^{h\nu/k T}	-1}.
$$

\noindent from the Planck function $B_\nu(T)$. Dropping the line source function below the Planck function in the line-formation layers tends to strengthen the line, while the line tends to become weaker in case of $S_{\nu}^{l}/B_\nu(T) > 1$.

\begin{figure*}  
	\centering
	\includegraphics[width=0.45\columnwidth,clip]{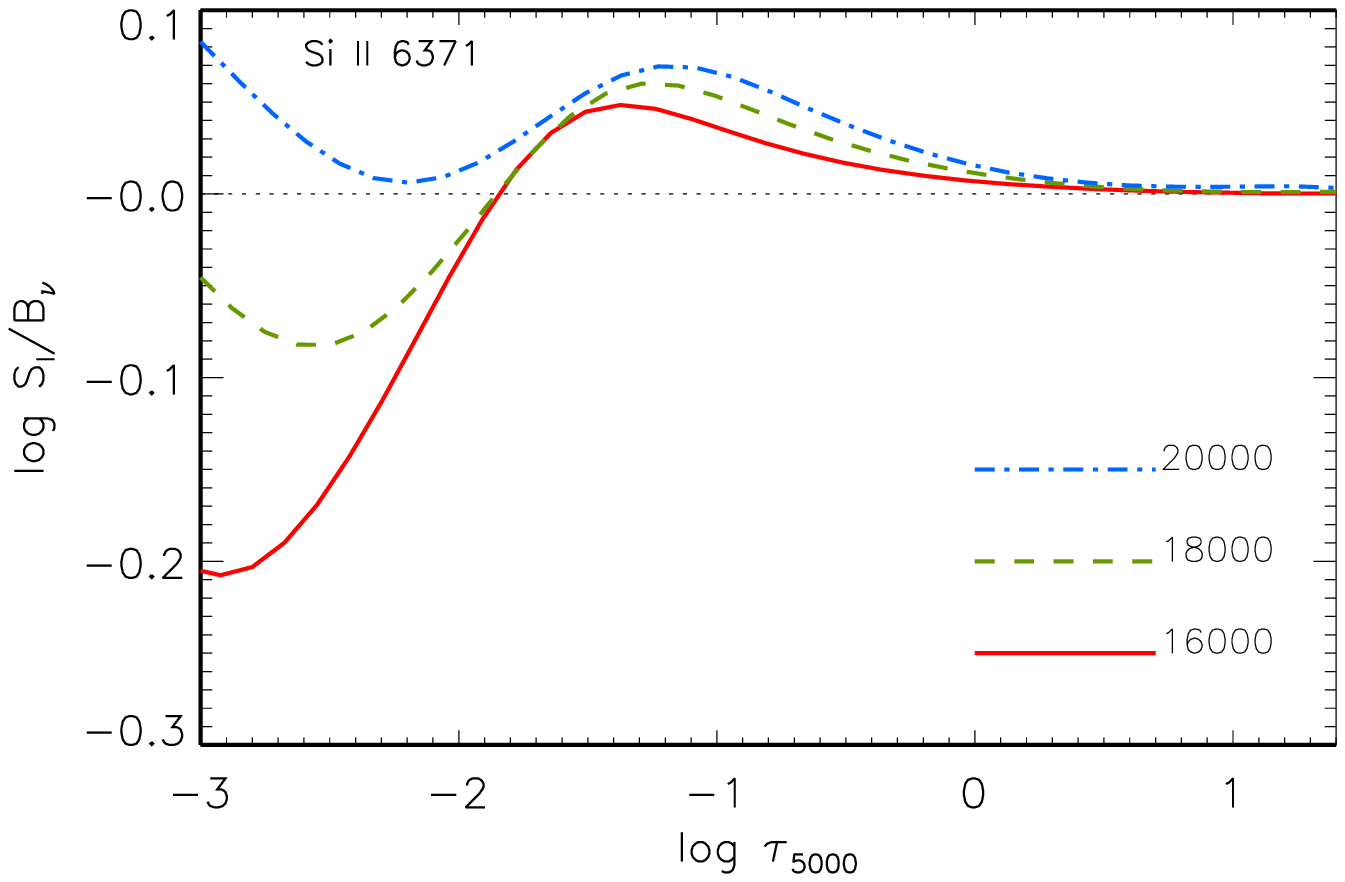}
	\includegraphics[width=0.45\columnwidth,clip]{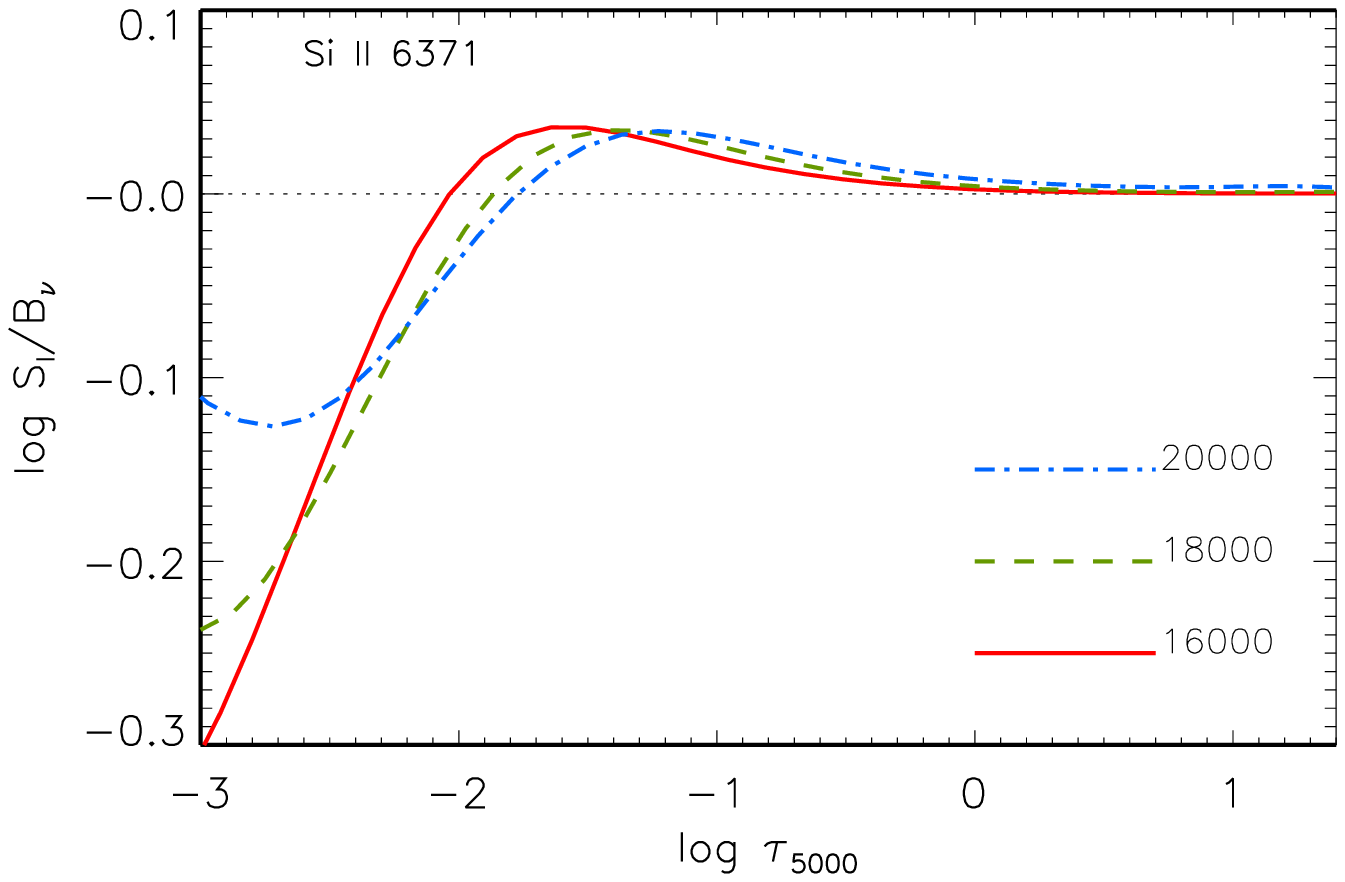}
	\caption{$S_{\nu}$(Si~II 6371~\AA)$/B_\nu$ ratio as a function of log~$\tau_{5000}$ in the model atmospheres of $\Teff$ = 16\,000, 18\,000, 20\,000~K and common $\logg$ = 4.0, [Fe/H] = 0, and [Si/Fe] = 0 from calculations with the model atoms Si2op (left panel) and Si2hyd (right panel).}
	\label{fig:sf}
\end{figure*}

Discrepancies in the departure coefficients between using Si2op and Si2hyd result in a rather  different behavior of $S_{\nu}^{l}/B_\nu(T)$ in these two cases. 
For a comparison with Takeda (2022), in our Fig.~\ref{fig:sf}, we plot $S_{\nu}^{l}/B_\nu(T)$ for Si~II 6371~\AA\ in the model atmospheres of $\Teff$ = 16\,000, 18\,000, 20\,000~K and common $\logg$ = 4.0 from calculations with Si2op and Si2hyd. For each of these model atmospheres, the differences in $S_{\nu}^{l}$ between Si2op and Si2hyd are, in particular, large for log~$\tau_{5000} < -2$. It is evident that our calculations with Si2hyd are very similar to that of Takeda (2022, Fig.~10, panel b).

\begin{figure*}  
	\centering
	\includegraphics[width=0.45\columnwidth,clip]{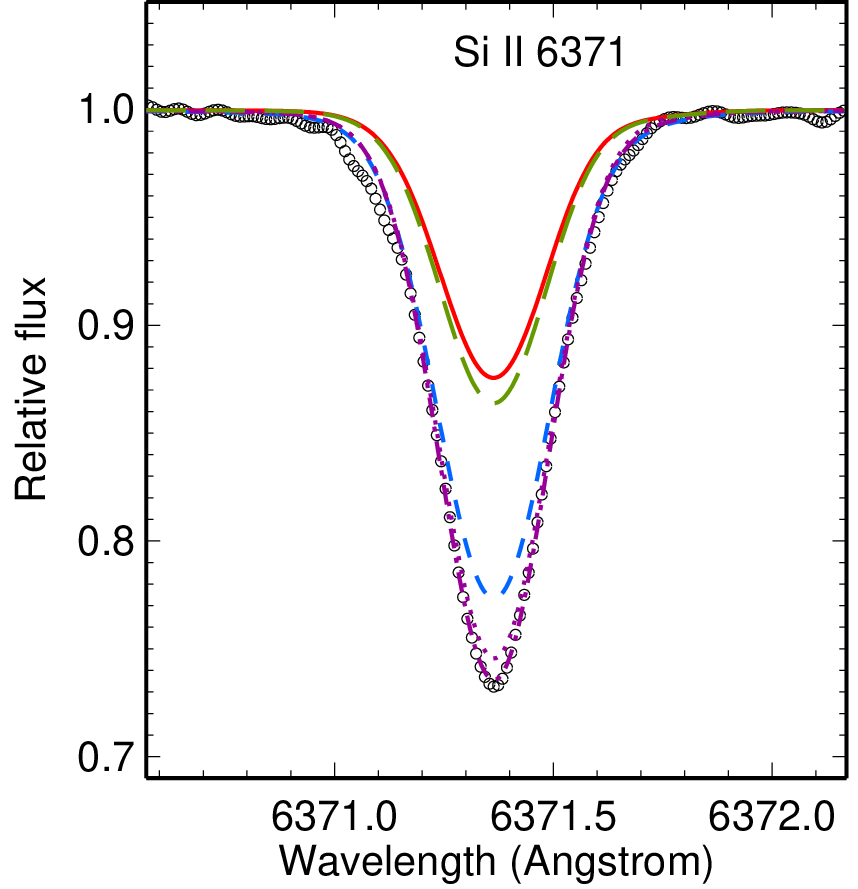}
	\includegraphics[width=0.45\columnwidth,clip]{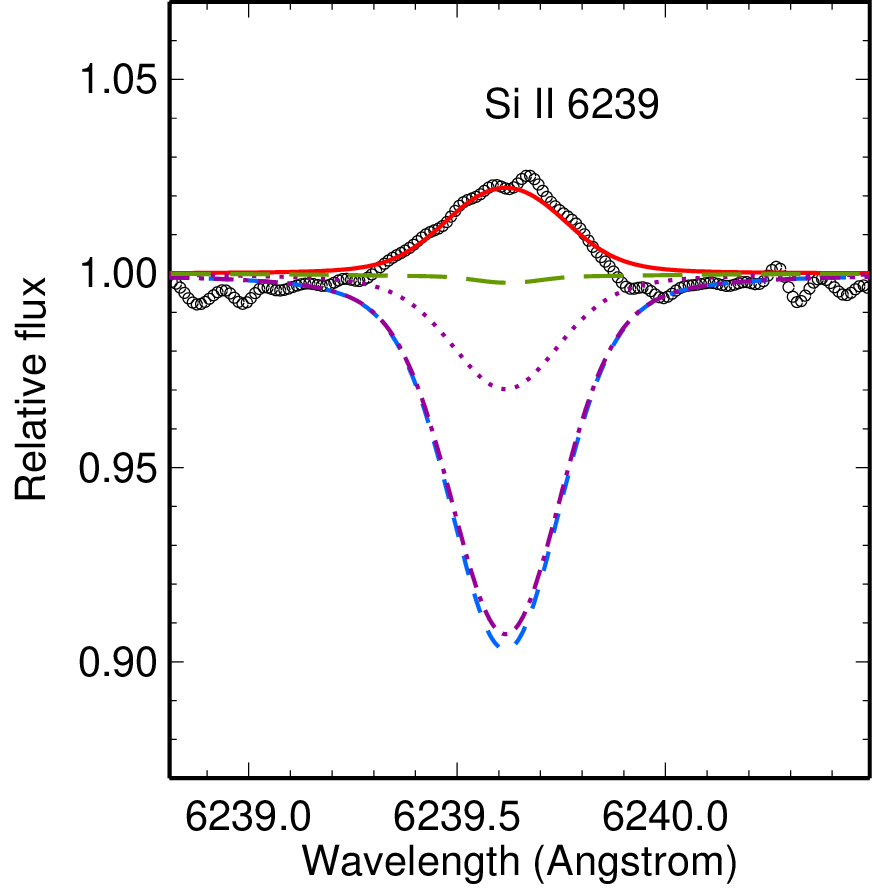}
	\caption{Line profiles of Si~II 6371~\AA\ (left panel) and 6239~\AA\ (right panel) in $\iota$~Her (open circles) compared with the theoretical profiles from the NLTE calculations using the model atoms Si2op (solid curve), Si2norad (long-dashed curve), Si2hyd (dash-dotted curve), Si2hyd3d4s (dotted curve) and the LTE calculations (short-dashed curve). The same abundance, $\eps{}$ = 7.41, was employed in calculations of Si~II 6371~\AA. For Si~II 6239~\AA, the best fit was achieved with $\eps{}$ = 7.92, while all the absorption profiles were computed with $\eps{}$ = 7.51.}
	\label{fig:lines}
\end{figure*}

\begin{figure}  
	\centering
	\includegraphics[width=0.88\columnwidth,clip]{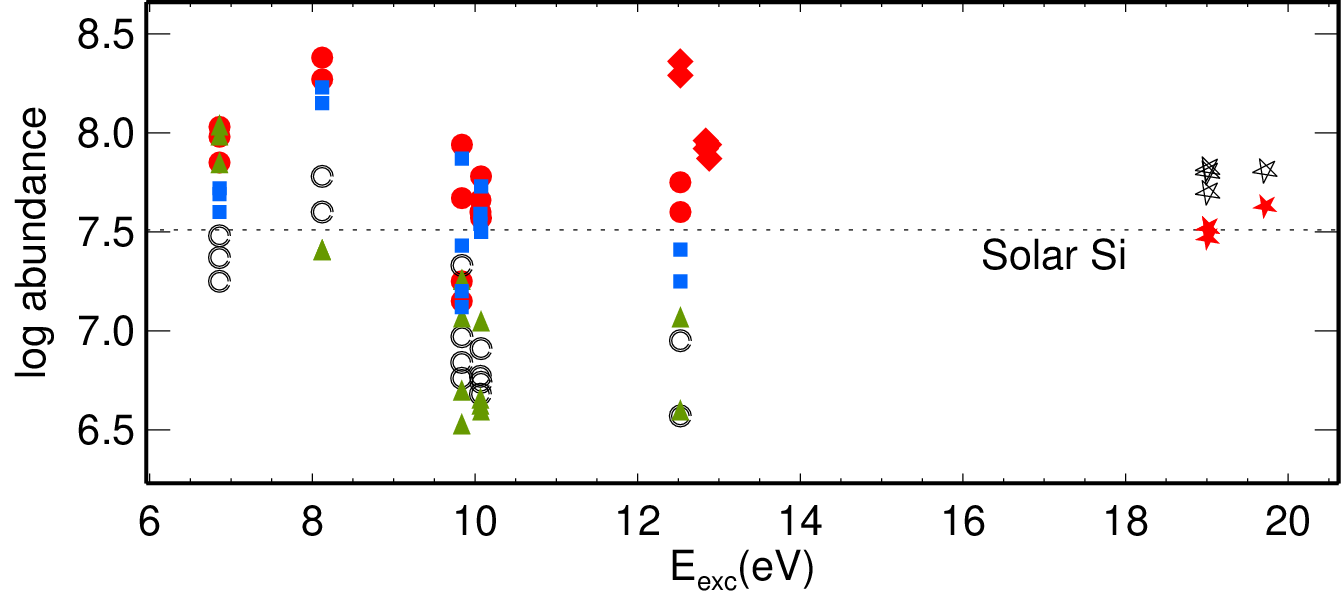}
	\caption{Abundances from individual lines of Si~II and Si~III in $\iota$~Her for various line-formation scenarios. Open symbols correspond to the LTE abundances. NLTE abundances of the Si~III lines are shown by filled five-point stars. For Si~II, the NLTE abundances from calculations with the model atom Si2op are shown by the filled circles (absorption lines) and the rhombi (emission lines); the squares are used in case of the Si2norad model atom and the triangles correspond to the Si2hyd model atom. The dotted line indicates $\eps{\odot}$ = 7.51$\pm$0.01 (Lodders 2021).}
	\label{fig:abund}
\end{figure}

The calculated $S_{\nu}^{l}/B_\nu(T)$ in the 17\,500/3.80 model atmosphere is very similar to that for $\Teff$ = 18\,000~K in Fig.~\ref{fig:sf}.
In case of Si2op, the Si~II 6371~\AA\ forms downwards log~$\tau_{5000} \simeq -1.8$, where $S_{\nu}^{l}/B_\nu(T) > 1$. Together with $b_{\rm low} < 1$, this leads to strongly weakened line (Fig.~\ref{fig:lines}) and a big positive NLTE abundance correction of $\Delta_{\rm NLTE}$ = 0.67~dex (Table~\ref{Tab:abund} and Fig.~\ref{fig:abund}). 

In case of Si2hyd, the core of Si~II 6371~\AA\ forms in the atmospheric layers with $S_{\nu}^{l}/B_\nu(T) < 1$. Strengthening the line core competes with weakening the line wings, resulting in stronger line compared with the LTE case (Fig.~\ref{fig:lines}) and negative $\Delta_{\rm NLTE} = -0.19$~dex (Table~\ref{Tab:abund} and Fig.~\ref{fig:abund}). Similar NLTE abundance correction of $\Delta_{\rm NLTE} \sim -0.1$~dex is reported by Takeda (2022) for Si~II 6371~\AA\ in the $\Teff$ = 17\,500~K model atmosphere.

\begin{table} 
	\caption{\label{Tab:abund} Silicon abundances of $\iota$~Her for various line-formation scenarios.}
	\centering
	\begin{tabular}{crrcccc}
\noalign{\smallskip}\hline\hline \noalign{\smallskip}
$\lambda$ & \Eexc & log $gf$ & \multicolumn{4}{c}{$\eps{}$}  \\
\cline{4-7}  \noalign{\smallskip}
 \ [\AA]  & [eV]  &          & LTE & Si2op &  Si2norad & Si2hyd \\
\noalign{\smallskip} \hline \noalign{\smallskip}
\multicolumn{7}{l}{Si~II, absorption lines } \\
3853.66 &  6.86 &  $-1$.34 &  7.25 &  7.85 &    7.60 &  7.85  \\
3856.02 &  6.86 &  $-0$.41 &  7.48 &  8.03 &    7.72 &  8.04  \\
3862.60 &  6.86 &  $-0$.76 &  7.37 &  7.98 &    7.69 &  7.99  \\
4075.45 &  9.84 &  $-1$.40 &  7.33 &  7.67 &    7.43 &  7.26  \\
4128.05 &  9.84 &     0.36 &  6.84 &  7.25 &    7.20 &  6.70  \\
4130.89 &  9.84 &     0.57 &  6.76 &  7.15 &    7.12 &  6.53  \\
4621.72 & 12.53 &  $-0$.43 &  6.95 &  7.60 &    7.41 &  7.07  \\
5041.02 &  9.84 &     0.03 &  6.97 &  7.94 &    7.87 &  7.07  \\
5055.98 & 10.07 &     0.52 &  6.68 &  7.66 &    7.59 &  6.63   \\
5056.32 & 10.07 &  $-0$.49 &  6.91 &  7.78 &    7.73 &  7.05   \\
5466.89 & 12.52 &  $-0$.06 &  6.57 &  7.75 &    7.25 &  6.60   \\
5957.56 & 10.07 &  $-0$.22 &  6.77 &  7.60 &    7.54 &  6.66   \\
5978.93 & 10.07 &     0.08 &  6.74 &  7.57 &    7.50 &  6.60  \\
\hline
\multicolumn{3}{l}{Average, 13 lines} & 7.07(0.41) & 7.74(0.25) & 7.56(0.23) & 7.08(0.55) \\
6347.11 &  8.12 &     0.15 &  7.78 &  8.38 &    8.23 &  7.41  \\
6371.37 &  8.12 &  $-0$.08 &  7.60 &  8.27 &    8.15 &  7.41  \\
\hline
\multicolumn{3}{l}{Average for Si~II} & 7.22(0.85) & 7.88(0.50) & 7.72(0.57) & 7.12(0.52) \\
\multicolumn{7}{l}{Si~II, emission lines}                      \\
6239.61 & 12.84 &     0.42 &  abs$^1$  &  7.92 &    8.83 &  abs    \\
6818.41 & 12.88 &  $-0$.52 &  abs  &  7.94 &    e+a$^2$  &  abs    \\
6829.80 & 12.88 &  $-1$.22 &  abs  &  7.87 &    e+a  &  abs    \\
7848.82 & 12.52 &     0.32 &  abs  &  8.36 &    8.82 &  abs    \\
7849.72 & 12.52 &     0.49 &  abs  &  8.29 &    8.73 &  abs   \\
9412.66 & 12.84 &     1.49 &  abs  &  7.96 &    8.76 &  abs   \\
\multicolumn{7}{l}{Lines of Si~III}                          \\
4552.62 & 19.02 &  0.29    &  7.82 &  7.47 &         &        \\
4567.84 & 19.02 &  0.07    &  7.80 &  7.52 &         &       \\
4574.76 & 19.02 & $-0$.41  &  7.70 &  7.52 &         &       \\
5739.73 & 19.72 & $-0$.10  &  7.81 &  7.63 &         &       \\
\cline{1-5}
\multicolumn{3}{l}{Average for Si~III} & 7.79(0.05) & 7.54(0.07) &  &  \\
\noalign{\smallskip} \hline \noalign{\smallskip}
\multicolumn{7}{l}{{\bf Notes.} The numbers in parentheses are the dispersions in the single line} \\ \multicolumn{7}{l}{measurements around the mean.} \\
\multicolumn{7}{l}{$^1$ absorption profile; $^2$ emission wings and absorption core. } \\
\end{tabular}
\end{table}
		
\underline{Si2norad versus Si2op.} Replacing the OP photoionization cross-sections with the NORAD ones for the Si~II levels leads to weaker NLTE effects for lines of Si~II. Although, with the Si2norad model atom, we cannot reduce the abundance discrepancy between Si~II 6347, 6371~\AA\ and the remaining absorption lines obtained by M2020 with Si2op. Even including these two lines in the average NLTE abundance, we obtain $\eps{SiII}$ = 7.72$\pm$0.57 (Table~\ref{Tab:abund}), which is closer to $\eps{SiIII}$ = 7.54$\pm$0.07 than in case of Si2op. Excluding Si~II 6347, 6371~\AA\ leads to consistent NLTE abundances from the two ionization stages.

However, in case of Si2norad, we fail to reproduce emission lines of Si~II (Fig.~\ref{fig:lines} for Si~II 6239~\AA). It is worth noting that, allowing the codes {\sc Synth}V\_NLTE + {\sc IDL binmag} to fit the emission line automatically, we can reproduce its emission profile, but the derived element abundance appears too high (Table~\ref{Tab:abund}), higher than the Si~III-based abundance by more than 1.2~dex. For high Si abundance, the line-formation depth shifts to the uppermost atmospheric layers with strong overionization of Si~II (see Fig.~\ref{fig:bf}) that favours the emission phenomenon for the lines arising between high-excitation levels.

\section{Summary and conclusions}\label{sect:conclusion}

This study shows that the SE of Si~II in the atmosphere of $\iota$~Her (B3~IV, 17\,500/3.80/0.02) is extremely sensitive to a variation in the Si~II photoionization cross-sections. We tested three sets of data, namely, equal accuracy photoionization cross-sections from the Opacity Project (model atom Si2op) and Nahar (1995) calculations (model atom Si2norad) and hydrogenic photoionization cross-sections (model atom Si2hyd). The NLTE abundances derived from the Si~II absorption lines with Si2op and Si2norad are different, by 0.18~dex, on average. With no one of the tested model atoms, we can fully solve the problems raised by Mashonkina (2020) in her NLTE analysis of Si~II-III in $\iota$~Her. We summarise pro and contra of the tested model atoms.

\noindent \underline{Si2op}

\noindent {\it pro:} reproduce 10 emission lines of Si~II in $\iota$~Her (M2020),

\noindent {\it contra:} obtain an abundance difference of 0.58~dex between Si~II 6347, 6371~\AA\ and the remaining absorption lines of Si~II, 

\noindent {\it contra:} obtain a supersolar abundance from Si~II 6347, 6371~\AA,

\noindent {\it contra:} obtain an abundance difference of 0.20~dex between Si~II (without 6347, 6371~\AA) and Si~III.

\noindent \underline{Si2norad}

\noindent {\it pro:} achieve consistent abundances from Si~II (without 6347, 6371~\AA) and Si~III,

\noindent {\it contra:} obtain an abundance difference of 0.63~dex between Si~II 6347, 6371~\AA\ and the remaining absorption lines of Si~II, 

\noindent {\it contra:} obtain a supersolar abundance from Si~II 6347, 6371~\AA,

\noindent {\it contra:} do not reproduce emission lines of Si~II in $\iota$~Her.

\noindent \underline{Si2hyd}

\noindent {\it pro:} achieve consistent abundances from Si~II 6347, 6371~\AA\ and Si~III,

\noindent {\it contra:} obtain an abundance difference of 0.33~dex between Si~II 6347, 6371~\AA\ and the remaining absorption lines of Si~II, 

\noindent {\it contra:} do not reproduce emission lines of Si~II in $\iota$~Her.

We conclude that, compared with the hydrogenic photoionization cross-sections, the data from Opacity Project provide a better treatment of the line formation for Si~II and Si~III lines in $\iota$~Her  despite, with applying this data, we calculate greatly overestimated NLTE effects for Si~II 6347, 6371~\AA. We call on atomic spectroscopists for further efforts to improve photoionization cross-sections for the Si~II levels. 

We find that, when using the hydrogenic photoionization cross-sections, we obtain the departure coefficients for the Si~II \eu{4s}{2}{S}{}{} level, the source function and the NLTE abundance correction for Si~II 6371~\AA, which are very similar to the corresponding values computed by Takeda (2022, Fig.~10). We suppose that the source of discrepancies between T2021 and M2020 lies with using different photoionization cross-sections. 

\acknowledgments
This study made use of the ESPaDOnS/CFHT archive, TOPbase, NORAD, and ADS\footnote{http://adsabs.harvard.edu/abstract\_service.html} databases. TS acknowledges support from the BASIS Foundation, Project No. 20-1-3-10-1.


\end{document}